\documentstyle[12pt]{article}

\begin{document}

\setcounter{page}{1}

\rightline{OSU--NT--94--09}
\vskip.5in

\centerline{\large {\bf Chiral Symmetry Breaking and Light-Front QCD}}
\vskip.3in
\centerline{ Kenneth G. Wilson and Martina Brisudova}
\centerline{ Physics Department, The Ohio State University}
\centerline{ Columbus, OH 43210}

\vskip.6 in
\centerline{\bf Abstract}
\vskip.1in
This paper consists of an overview of the discussions on chiral symmetry
breaking followed by a transcript of the discussions  themselves.
\vskip.1in

\vskip.3in
{\bf 1. Introduction}
\vskip.1in
This article covers discussion sections on chiral symmetry breaking.

First, let us outline  the essential questions which
must be answered on the way to solving QCD as a few body problem.
\vskip.1in
The first thing one needs is the canonical Hamiltonian. The canonical
Hamiltonian is well defined except for those peculiar modes which have
precisely zero longitudinal
momentum, so called zero modes.
In the exact light-front (LF) theory these would
surely give rise to a nontrivial vacuum and
have a key role in both confinement and
 spontaneous chiral symmetry breaking. Here we wish to remove these
states from the theory but argue that additional
 effective interactions in the Hamiltonian can restore the same physics.

There are several ways of removing these states from the theory. The
simplest one is to introduce a cutoff $\epsilon$ such that
 all $k^+ > \epsilon$.
Then one has to introduce counterterms to subtract infrared divergences.
The finite parts for these counterterms would be a logical source for the
necessary additional terms.
Another possibility is to truncate the theory in
a finite volume as in DLCQ. In this case one must face the problem of
``constrained" zero modes. We prefer the first approach because we do not
know how to solve the constrained mode problem.
\vskip.1in

Next question on the list regards the cutoffs
used to regulate the theory.
One could use cutoffs that either maintain the separation between
longitudinal and transverse degrees of freedom or ones that mix them
(for more detailed discussion see contribution by R. Perry).
\vskip.1in

A fundamental feature of QCD is that it is a confining theory. Any attempt
to solve for QCD bound states must implement this property, which is believed
to be related to the nontrivial QCD vacuum. If we try to describe QCD bound
states as a few particle states, can we identify or implement
a confining mechanism
which does not require infinitely many wee partons?
\vskip.1in

The answer to this question is: Yes.
As opposed to our earlier work [1],
the confining mechanism no longer requires an artificial potential.
If one assumes that the gluon is
massive, then the divergent parts of the instantaneous interaction and one
gluon exchange do not cancel entirely.
A mechanism for obtaining logarithmic confinement due to an incomplete
cancellation has been obtained by Perry. Perry's confinement mechanism is
independent of the initial cutoff used.
I have oversimplified his
argument in proposing a $k^+$-dependent gluon mass (see my opening talk).
\vskip.1in

An exact treatment may require new effective interactions as well as the
gluon mass. But a gluon mass is sufficient to get calculations started. It
cannot be ruled out because a gluon mass can be understood as a finite part
of an infinite renormalization which is needed anyway.
\vskip.1in

The assumption that the gluon is massive violates gauge invariance, but it is
not unnatural. Apart from the fact that a similar result may be
emerging from the zero
mode analysis (see Pinsky's comment below and the contribution of
A. Kaloniatis ),
one can get a hint from the hadronic spectrum.
If gluons were massless like photons, the spectra would be continuous
rather than discrete.
\vskip.1in

\vskip.1in

The main subject of these two discussion sections is contained in
the question:
Where does chiral symmetry breaking come from?

It is important to realize that light front chirality is not the same as
equal time chirality. On the light front chirality coincides with
helicity.
It is important to remember that chirality is typically broken only by
terms linear in $m_q$. The light front
free Hamiltonian does not violate chirality on
the light front, because it involves only $m_q^2$.

In the absence of the zero modes, we will argue that
 the spontaneous chiral symmetry breaking has to be realized through
explicit symmetry breaking interactions.
So there would be two types of explicit symmetry breaking on the light front -
one corresponding to a nonzero pion mass and one reflecting the
spontaneous symmetry breaking.

As opposed to the confining mechanism, which can be obtained as an
incomplete cancelation, chiral symmetry breaking requires adding a new
interaction. The incomplete cancelation causing confinement
preserves the chiral symmetry
of the canonical theory.

For an illustration of how complicated the issue of chiral symmetry breaking
is let us
consider the canonical Hamiltonian including the only
explicitly chiral symmetry breaking term
(the spin flip part of the one
gluon exchange). If one drops the explicit breaking term,{\footnote{Strictly
speaking, this is not a legitimate operation, as
explained in more details in the discussion. Linear and
quadratic mass should be linked. We make this operation for the purpose
of ``playing" with the theory. It enables us to demonstrate clearly a
fundamental reason why it is necessary to introduce new symmetry breaking
interactions.}}
entire $SU(3) \times SU(3)$ multiplets will have the same mass. Therefore,
to make the pion massless and the $\rho $ heavy without the explicit
breaking, one has to add a new
interaction. The precise operator form and the coefficients can, in
principal, be
determined from the requirement that in the chiral limit the chiral current is
conserved. (There are however
some subtleties regarding renormalization of the $-$ and $\perp $ components
of the current (see contribution of A. Harindranath.)).

To understand what the structure of the new terms might be we reason
by analogy with the sigma model. However, there are some
important differences between the symmetry  breaking mechanism in the
$\sigma $ model versus QCD.

For a scalar $\sigma $
 field, the zero mode of the $\sigma $ is shifted by a constant
when there is a spontaneous symmetry breaking.
 Then this constant
generates new terms when substituted back into the canonical Hamiltonian
which involve the zero mode of $\sigma $.
 In QCD, it is  $\bar{\psi} \psi $ that has a
nonzero vacuum expectation value, but a constant shift for
$\bar{\psi} \psi $  has no effect. However, we do already substitute a
formula for the lower components $\psi ^{lower}$ to derive the canonical
Hamiltonian. This provides an opportunity for changes to the
zero mode of
$\psi ^{lower}$ to occur before making the substitution.
More details will be provided in the next section.

\vskip.3in
Finally, I will comment on the picture just presented.
Some parts of the program described will be less convincing than others, and
 unexpected obstacles, which we do not foresee yet,
may still arise. This is
typical at the beginning of a scientific
revolution in Kuhn's sense, even a modest one.
 You may accept the new approach now, when the revolution is just past its
diaper stage, or you may wait.
But, as Prof. Susskind said in
this discussion: {\bf``If you wait till he can convince you, it'll be too
late."
}

\vskip.3in
{\bf 2. Transcript of the discussions on chiral symmetry breaking}
\vskip.1in

\vskip.1in

{\bf Wilson:} We have had a lot of discussions and so forth, but I am sure
there
are many of you who still don't have  a picture of what it is that the OSU
group is really doing here.

Well, that just puts you in a situation that we were in for 3 years until
sometime this spring. And the first thing I want to say is that  we would
have never gotten through that period of confusion, in fact we would have
never get started on it, except for our host here, Stan G\l azek, who came
to OSU and said: We had to accept the idea that the proton is 3 quarks. We
could not engage in
fuzzy thinking about lots of gluons, we had to
think of it as of 3 quarks and we had to answer all the questions about
those 3 quarks. We have to answer why they are confined and we have to answer
why chiral symmetry is spontaneously broken and we can't make the excuse
there are lots of gluons and quark - antiquarks so we don't know what's
happening. And Stan just hounded us.

There was no way we could do this except
by working through all the pieces of the problem, and not just
pieces
A B C, but
pieces
A B C D E F G H I J K L M N, until we went through
every problem about 3 times each that we
could start putting it together and we realized that there is a sensible
picture.

So all I can do is to write down what are the labels for ``A B C D E F G H"
 so that
you understand why we went crazy.

\vskip.1in

\noindent {\bf A} The first thing you have to have is a {\bf canonical
Hamiltonian}. It is not that there is any real argument about what the
canonical Hamiltonian is, except  when it  comes to the zero mode. You have
that question about how {\bf zero modes} change the canonical Hamiltonian
itself.
\vskip.1in

\noindent {\bf B} You need to cut it off. How do you {\bf cut off?} Do we
use a cutoff that maintains the separation between longitudinal and
transverse directions or do we mix them?
\vskip.1in

\noindent {\bf C} Can we get {\bf confinement without an infinite number
of wee partons?} The whole picture formerly was that you have to have an
infinite number of wee partons
to cause confinement, in other words,
to reflect the nontrivial vacuum of normal rest frames.
\vskip.1in

\noindent {\bf D} Where does {\bf chiral symmetry breaking } come from?
\vskip.1in

\noindent {\bf E} {\bf How do you renormalize?}
What is the situation when the QCD Hamiltonian has a big cutoff and how do
you relate it to a qualitative picture with a small cutoff? How to ensure
the physics is independent of the big cutoff?
\vskip.1in

\noindent {\bf F} {\bf What are the parameters?}
You face the problem that when you set things up in the LF theory, all of a
sudden there are lots of parameters that you can throw
into the canonical Hamiltonian
because you do
not have the full Lorentz invariance to control the number of parameters.
Just look at the canonical Hamiltonian: unless you can connect the mass
term that appears in the free Hamiltonian to the mass terms that appear
throughout the interaction, you suddenly find yourself with lots of
parameters and as soon as you renormalize, it looks like those parameters
disconnect from each other. So this was the
problem
that lead to the coupling
coherence idea.
\vskip.1in

\vskip.3in
{\bf Question (Van de Sande):} What is a one word answer to C?

{\bf Wilson:} The answer is yes, if you accept gluon $k^+$
 dependent mass. That is
there is a way of constructing the gluon mass where the mass squared is
proportional to $k^+$ (not a constant, the way
it was proposed in our paper) where you
get the log potential that we are talking about as a confining potential in
the second order, independent on how you do the cutoff. So the answer to
this question is yes. That does not say it is physically correct, it
simply says:
Yes, we can get confinement out of the canonical Hamiltonian
without an infinite number of wee partons.

\vskip.3in
{\bf Question (Susskind):} Can you get a linear potential?

{\bf Wilson:}
In the second order of perturbation  theory, we get a logarithmic
 potential. A sum to all orders might still give a linear potential. We do
not know what happens beyond second order.

\vskip.3in
{\bf Question (Pauli):} How do you calculate the gluon mass?

{\bf Wilson:} At the ABC level it is a parameter.
I agree that in the complete theory you can't have a gluon mass that is a free
parameter but at this ABC level you can not calculate the gluon mass.

{\bf Perry:} If you don't like to put in the
gluon mass by hand, you can also do the second
order calculation with cutoffs and find out what kind of masses you have to
put in (i.e. as counterterms).

\vskip.3in
{\bf Question (Susskind):} Could it be that these are somewhat dependent on the
scheme of renormalization?

{\bf Wilson:} Let me qualify this:
The question C is, can we get confinement in the limit all
cutoffs removed?
Thus the answer should not be cutoff dependent. It should not
be scheme dependent.
The only case that would forbid our $k^+$ dependent term would be a
renormalization scheme that
a) preserves longitudinal boost invariance and
b) avoids counterterms with an explicit dependence on the total center of
mass momentum $P^+$. To my knowledge no renormalization scheme meets this
requirement.

\vskip.3in

{\bf Pinsky:} Let me just illustrate how I see a completely different way up to
the valley. We saw in Alex'es talk that the topological structure in
$1\over{(k^+)^2}$ interaction had an additional piece which came from
topology and looked  like the  mass of the gluon. So when I see something
like that I say maybe the right path is to try to see the connection
between the topology of the instantons.
It has lots of familiarity with what you are saying but there is a
difference between this and the self consistent treatment (put in the mass
and show that it works).

{\bf Pauli:}
It seems to me a lot
more ad hoc to put in the gluon mass by hand.

{\bf Wilson:} My statement that you can get anything you want out of the
 zero modes is
clearly extreme and unwarranted. What I would say about the zero modes is
that from everything I can see it will take you a long time to resolve
the problems of the zero modes.
Meanwhile I will be working with more ad hoc approaches to the theory
without zero modes.
But it is the only thing I can say about it.

{\bf G\l azek:} I understand that zero modes are not necessarily unimportant. I
don't agree that we know that
they will never be important. They may be important in
ways we don't know yet. When I was thinking about LF QCD I wanted to
include those modes. Actually, I succeeded in designing some zero mode
rules that could reproduce some QCD sum rules and I was excited about it.
I thought that it was the most important part of LF QCD that we should work
on. But when I started to use the Hamiltonian that I designed to reproduce
QCD sum rules I discovered that I have difficulties that  cannot overcome
problems that completely annihilate the possibility to use this Hamiltonian
in a very  well defined procedure of  calculating numbers for hadrons.
It went to the extent that I was able to build a very simple model of
mesons and hadrons by analogy with the QCD Hamiltonian that reproduced
Carl-Isgur constituent model parameters within 10 \%. But that model had
literally nothing to do with QCD, because when I started with QCD it was
impossible to do anything unless I knew how to renormalize.
Before we'll be able to answer questions concerning the vacuum,
zero modes and things
like that, at least to me it became clear that first we have to
learn how to do renormalization.

{\bf Wilson:} Let me say a few words about the zero modes. I apologize for the
attitude I take toward zero modes which is sometimes hard for me to control.
But I perfectly agree with what Stan says that very important things
may come out of the zero modes analysis, things to which we may not
get to any other way, but at the present time, the only way I have to get
through this cycle of all the ABCDEF you have to understand to make the
picture, the zero
modes are not part of that cycle and I am still optimistic that we can make
progress whether or not the insights of the zero modes that we need are
forthcoming or not.

\vskip.3in
{\bf Question (Susskind): } I want to ask my question again, because I
am not convinced that I believe your answer. This was the question about
the scheme dependence in the effective potential. You are finding these
effects, that is chiral symmetry
breaking and confinement, due to an  incomplete cancelation of contact
terms at very low momentum expansions. Is that right?

{\bf Wilson:} No, the chiral symmetry breaking is not an incomplete
cancelation. The chiral symmetry breaking term will be bringing in a new
concept, a new structure into the canonical theory, that I will claim could
come from the zero modes. It will be a new structure which you can't create
by any incomplete cancelation, because any kind of cancelation would
preserve the chiral symmetry
of the canonical theory and I have to put in something that breaks the
chiral symmetry. I am saying: to get the confinement we didn't
have to put in anything, it came from an incomplete cancelation of what was
already there, but for chiral symmetry we have to put in something.

{\bf Question (Susskind): }... and have an incomplete cancelation?

{\bf Wilson:} No. It's just the question of putting in something.

\vskip.3in

CHIRAL SYMMETRY BREAKING.
\vskip.3in
{\bf Wilson:} I want to do a bit of set up on chiral symmetry. I want to
show how desperate we became before I show you the finesse.
I want to start with the canonical light front Hamiltonian, no zero modes,
and I am not going to write out everything, I'll just remind you that there is
a free part where the quark energy is
\begin{eqnarray}
{p_{\perp}^2 +m_q^2 \over{p^+}}
\end{eqnarray}
 and
an analogous gluon term.
Then there are the quark-gluon interactions,
one term of which is the explicit chiral symmetry breaking term of the form
\begin{eqnarray}
 g m_q \bar{\psi} \sigma _{\perp} \cdot A_{\perp} \psi
 \end{eqnarray}
where $A_{\perp}$ represents the gluon and $\sigma _{\perp}$
 are the Pauli matrices in transverse light front coordinates.
Color and flavor indices have been suppressed.
Among all other terms we will be interested only in the instantaneous quark
exchange term - a two quark-two gluon interaction. For example, in this
term a fermion and a gluon come in and a fermion and a gluon come out, what
you diagrammatically represent as a quark exchange even though it's not an
actual particle exchange diagram, just a term in the canonical Hamiltonian.

\vskip.1in
If $g=0$, so that  all what we have are the free terms, and no zero modes,
I remind you that what we have at this stage is an exact $SU(6)$ symmetry
of all the 3 flavors. That is assuming that the 3 quarks have the same
nonzero
mass. Of course the two spins (the two helicities of the
quarks) have the same mass. That just simply means that you have a
trivial $SU(6)$ symmetry.
\vskip.1in
In the presence of the interaction, $SU(3) \times SU(3)$ symmetry is
maintained
if term $(2)$ is  dropped
out.
Term $(2)$ is the only term that can
 break explicit $SU(3) \times SU(3)$, so if I
turn on the interaction but not this term, I continue to have full
$SU(3) \times SU(3)$ multiplets.
(The larger $SU(6)$ symmetry is now broken by terms with
$(\sigma _\perp \cdot A_{\perp})^2$ type chiral structures.)
 Of course funny things
can happen to the masses due to these interactions
but funny things cannot happen within a given
$SU(3) \times SU(3)$ multiplet. The $SU(3) \times SU(3)$ multiplets  have
to always maintain the same mass through out all members of the multiplet.
So if I send the $\pi$ mass to zero, without the term $(2)$ the
entire multiplet goes to zero. This does not make any sense.
To resolve the situation I need to things. First I want to decouple $(1)$
from $(2)$.
We consider what happens if $g$ is nonzero but the mass $m_q$ in $(2)$
is  a separate parameter $m'_q$ from the free quark mass $m_q$. What I have
in mind is that the free mass should be a constituent mass $\sim \ 300 $
MeV, while $m'_q$ would be much smaller or zero, like a current mass. In
the true relativistic theory the values of $m_q$ and $m'_q$ would be
linked, but for now I will ignore the linkage.

Secondly, we still need an extra term in
 the Hamiltonian as a result of subtraction or
infrared behaviour or something else,
which does the job of supplying another explicit $SU(3) \times SU(3)$
symmetry breaking
 which involves constituent masses as coefficients rather than current
masses and therefore provides a large separation of the $\pi$ and $\rho$
masses. This term should arise as a result of eliminating zero modes from
the theory and should maintain local chiral current conservation - the true
signal of spontaneous breaking only.
\vskip.1in
{\bf Perry:} When you throw away the zero modes and look at how the
parameters renormalize, you will see that the quadratic mass has an infrared
divergence. The linear mass has no such divergence. So they do renormalize
quite differently.
\vskip.1in
{\bf Question (Susskind):} Does it have  to do with anything else
than you just have to renormalize everything separately unless they are
linked by a symmetry?

{\bf Wilson:} Right, but it's more complicated because you're getting to
the coupling coherence which Bob (i.e. Perry) has mentioned but not really
talked about.
\vskip.1in
{\bf Question (Burkardt):} This mass (i.e. term $(2)$) is determined by the
pion mass scale. To get the $\rho $ heavy means you have to add another term,
 right?

{\bf Wilson:} Yes. That's the whole point that I am saying. We
are missing something that would allow this (i.e. the linear
mass in term $(2)$) to stay
at the pion mass scale, this thing
(i.e. quadratic mass in term $(1)$) to be at $\rho $ mass scale and yet
still not have the $SU(3) \times SU(3) $ near symmetry
driving everything down to zero.
\vskip.3in
{\bf Question (Berera):} Ultimately, there should be one $m_q$.

{\bf Wilson:} The statement that there should be one $m_q$ comes back to
what Bob Perry has worked out called coupling coherence.
When you look at the renormalization process to higher than to second
order, (you have to go at minimum to third order) you find out that all
the appearances of $m_q$ are linked exactly the way that you would like.
{\footnote { For more detailed explanation of coupling coherence
see R. Perry's paper in this volume.}
\vskip.1in
{\bf Question (Susskind):} But they don't come up the same way, do they?

{\bf Wilson:} They don't come up the same, but they are linked.
\vskip.1in

{\bf Question (Berera):} In your language, does it basically mean that
 there is one direction
 irrelevant? If you start with two dimensional space and there is only one
parameter at the end ...

{\bf Wilson:} No, no. It's not a question of relevant versus irrelevant. It
is a question of preservation of  the full Lorentz symmetry of the theory.
The preservation of the full Lorentz invariance translates into something
called coupling coherence, that forces you to set up linkages between all
the parameters. If you break the linkage by letting $m'_q$ be arbitrary all
that happens is that Lorentz invariance is lost.

{\bf Question (Berera):}
So it has nothing to do with the renormalization flow?

{\bf Wilson:} It has to do with the renormalization flow but the criteria
that we put on say that the renormalization flow should be describable in
terms of one coupling constant rather than an infinite number of coupling
constants all doing their own thing.
\vskip.1in
{\bf Question (Susskind):}  Is it anything like what would happen in the
Hamiltonian lattice gauge theory when you allow different coefficients in
front of spatial and time derivatives?

{\bf Wilson:} That's right.
\vskip.1in
{\bf Question (Berera):} I am glad you just made the distinction because when I
think about the relevance and irrelevance I think about it from dynamical
point of view. Are you telling me that I am going to get one coupling, I
mean one parameter, from the
kinematic structure?

{\bf Wilson:} It's not a kinematic structure. Lorentz invariance is not
explicit in this theory, so it is a very subtle dynamical issue to keep
those parameters linked. You have to look at the details of renormalization
to see what coupling coherence means.
\vskip.1in
{\bf Question (Berera):} How does it relate to your old picture of relevant
and irrelevant operators?

{\bf Wilson:}
It is not a question of relevance versus irrelevance. Both operators are
relevant - the free mass term and the chiral symmetry breaking interaction.
There is however a unique linear combination which preserves Lorentz
invariance. Any other combination is still relevant, but would destroy
Lorentz invariance.

Similarly, there are many marginal operators independent of the mass in the
canonical Hamiltonian. But again only one unique linear combination
preserves Lorentz invariance. All other combinations remain marginal but
violate Lorentz invariance.
\vskip.3in
AFTER THE LUNCH the discussion continued addressing the question how one
can find new chiral symmetry breaking terms.
\vskip.1in

{\bf Wilson:} What I am going to do is I'll start with the $\sigma $ model
but I am not going to do any development. I'll put the chiral symmetry
breaking in the $\sigma $ model into one equation and then use that as
 a starting point for discussion.

\vskip.1in
{\bf Question (Yan):} Are we talking about explicit or spontaneous
symmetry breaking?

{\bf Wilson:} What I need is an explicit symmetry breaking term, but it is
supposed to be representing the spontaneous.
There are two different explicit symmetry breaking terms in this
formalism.
\vskip.1in
I remind you,
the way we handle the $\sigma $ model{\footnote{the linear sigma model}}
 is that we translate the sigma field,
but I just want to write that translation in a particular way. I want to
take a term in the Hamiltonian, say,
\begin{eqnarray}
\int d^3 x \ \phi ^2(x) \sigma ^2 (x)
\end{eqnarray}
before translating. And before translating I am going to rewrite this in
the momentum space in terms of the Fourier transforms:
\begin{eqnarray}
\int d^3 k \int d^3 k_1 \int d^3 k_2 \phi _{k} \phi _{k_1}
\sigma _{k_2} \sigma _{-k -k_1 -k_2}
\end{eqnarray}
where $k$, $k_1$ and $k_2$ are light cone three vectors.

And now I am going to look at the effect of translation on this term, an
absolutely trivial exercise except that I am going to do it in the momentum
space. I am going to take the $\sigma _{k_2}$ and look at
a translation in momentum space form:
\begin{eqnarray}
\sigma _{k_2} \ \rightarrow \ \sigma ^{'} _{k_2} + \delta ^3 (k_2) \cdot C
\end{eqnarray}
where $C$ is a constant which corresponds to the vacuum expectation of
$\sigma $. In other words, when you have the condensate (to use the
word that everybody likes to use here) and you have to look at it in the
momentum space, that corresponds to a delta function added to the field
 $\sigma $.
\vskip.1in
In a diagram context, if I have a four point vertex where two of the fields
are $\phi $ and two of the fields are $\sigma $, what
this gains is a term
that has three external
fields and a fourth leg which has an x on it (i.e.
an external field contracted with the
vacuum) is the fourth leg pinned at $k=0$.
That is all I want to say about the $\sigma $ model.
\vskip.1in
The idea here is that any time you have  a $\sigma $  line, you can
replace it
with a constant $C$ and pinning its momentum to zero. We have no a priori
information of what the constant $C$ is and   we don't even know
whether  we should put
the same constant in two locations.

{\bf Susskind:} In particular, it does not have to be a classical constant.
\vskip.1in
{\bf Question (Namys\l owski):} But when you pin the leg to $k=0$, isn't
it  the zero mode?

{\bf Wilson:} Yes, the pinning
is obviously connected to zero mode, the mode with
all components of $k$ vector equal
to zero, so the only way you can properly justify this substitution
is through the zero
mode theory.

\vskip.1in
{\bf Question (Susskind):} Is the rule of the game: We don't know what the
value of the coefficients is but we adjust it so that the pion is
massless?

{\bf Wilson:} Yes, but
this adjustment must do more than give a
zero pion mass. What you want is to ensure a locally conserved
chiral current
when the pion mass is zero.
By insisting that you have conserved current you can derive
the standard shifted Hamiltonian
of the $\sigma $ model
without ever explicitly invoking information about zero modes.{\footnote{
There are some subtleties
 regarding the $-$ and $\perp$ components of the current.}}
Detailed analysis can
be found in the appendix of our paper [1].
\vskip.3in
{\bf Wilson:}
Now turn to QCD. What we are demanding is a nonzero expectation value for
$\bar{\psi} \psi$ instead of an elementary field. So you go around diagrams
and say: Is there any way that I can arrange to have  a constant vacuum
expectation value for $\bar{\psi} \psi$ change
 the canonical Hamiltonian?
And as you can well
imagine, we could not produce a changed Hamiltonian. There was no way
that you could shift the $\bar{\psi}\psi$ and have it mean anything in the
context of the canonical QCD Hamiltonian.
So now we come to the stretch.
\vskip.1in
{\bf Question (Burkardt):} Could not you contract through one of those
instantaneous fermion lines?

{\bf Wilson:} You have found the right place. But the way I want to
motivate this is I want to look back at how the instantaneous fermion
term arises.
I haven't taken you through the canonical theory, but in any case,
you had in Harindranath's talk this morning the relevant equation, the equation
which solves for the lower components of the $\psi$ field.
\vskip.1in
The
instantaneous quark exchange term originates
with an ordinary
 three point vertex, but with one quark leg referring to a lower component.
But the lower component is not an independent degree of freedom and it
has to be eliminated. It gets
replaced by the following structure - somewhat simplified:
\begin{eqnarray}
\psi _q ^{lower} = {1\over{q^+}} \int _p \sigma _{\perp} \cdot A_{\perp \ p}
\psi _{q-p} ^{upper} +{1\over{q^+}} m \psi _q ^{upper}
\end{eqnarray}
Now I want to add a term to this structure reflecting the effect of the zero
modes, because
the picture is that the whole physics of chiral symmetry breaking has
to come from zero modes. That is the only place where you can have
nontrivial vacuum.
\vskip.1in
 I have to tell you that this is different
from the presentation
 in our paper. I worked out this presentation just for this conference.
\vskip.1in
People can make all sorts of pictures of how the zero modes might affect
the chiral symmetry. But I try to copy what was done in the
$\sigma $ model. So I want to add something that is associated with
momentum zero. Let me show you an example of
 what I add in:{\footnote{ This is a
 slight modification of the expression Dr. Wilson
wrote down first, which was done later in the discussion
thanks to a comment by Matthias Burkardt.}
\begin{eqnarray}
 + \ \delta (q^+) \int _{q'_\perp} \vec{f}_{\perp}
(q_{\perp} - q'_{\perp}) \cdot
\int A_{\perp \ p} \psi^{upper} _{q' - p} d^3 p
\end{eqnarray}
where $q'$ is understood to differ from $q$ only in its
transverse components. Now,  notice
that I do not force the  momenta $q_{\perp}$ to be zero.
Furthermore, when you are taking an
expectation value of $\bar{\psi}\psi$ you don't demand that both fields
have zero momentum, only that the product has a zero momentum.
Hence my rule is that the $A_{\perp}$ and $\psi$ can separately be nonzero
modes, as long as
the product lives in
in the zero mode sector.
The form factor $\vec{f}_{\perp}$
reflects the properties of the vacuum. The vacuum knows about the masses,
so $\vec{f}_{\perp}$ can have very complex mass
dependence, rather than a structure dictated by power counting in the
absence of mass dependence
\vskip.1in
I am not going to justify the precise term I wrote here. I am saying it is
reasonable to consider. You may consider some other terms.
\vskip.1in
{\bf Question (Soldati):} But you can have only $q^+ \not= 0$ in an
exchange.

{\bf Wilson:} No, in the instantaneous diagram $q^+$ can be anything. It
just depends on whether the incoming momentum of the fermion is bigger,
smaller or equal to the momentum of the outcoming fermion. And three are
possible. But what I am going to say is that for the added term
to be associated with
the zero modes, I have to put in a delta function that forces $q^+$ to be
zero.

\vskip.1in
{\bf Question (Namys\l owski):} But at $q^+ =0$, wouldn't that blow up?
(the first two terms.)

{\bf Wilson:}
No, the first two terms remain functions of $q^+$, with $q^+$ not forced to
be zero. However, if quarks and gluons
actually existed as asymptotic particles, there is no
way that I could have the
delta function. That would generate a physical nonsensical quark-gluon
scattering amplitude.
 What that says is that the
kind of things I am doing you can only do in a confined theory. Because in
the confined theory you can never have quarks or gluons
as free particles which means
they will only appear as constituents in a wave function. That's the only
way they will have a physical meaning.

{\bf Question (Namys\l owski):} But even being confined the instantaneous
momentum can be equal to zero.

{\bf Wilson:} That's right.

{\bf Namys\l owski:} Oh, they will be hidden in a hadron.

{\bf Wilson:} Yes, every one of the quark momenta will
be hidden in a hadronic wave function so there is always an integration
over these momenta to compute physical amplitudes.
But there is nothing to stop the momentum of an incoming fermion to be equal
to the momentum of the outcoming fermion as part of the integration.
And of course, since there is an integration, there is no problem with the
delta function.
\vskip.1in
{\bf Question (Burkardt):} Am I right in assuming that after inserting this
extra term into the  Hamiltonian all external legs are at nonzero $p^+$?

{\bf Wilson:} Yes, you still keep the infrared $\epsilon$ cutoff on the
external legs. How I handle any cutoff in the instantaneous exchange is not
necessarily linked to what I do on the external legs.
\vskip.3in
{\bf Question:} Is the mass in eqn. $(6)$ a constituent or current mass?

{\bf Wilson:} The mass in eqn. $(6)$ starts out as a current mass
in the free field limit, but it
will renormalize and it may do all sorts of crazy things when you
renormalize.

{\bf Question (Namys\l owski):} But it cannot reappear as a constituent
mass, can it?

{\bf Wilson:} I will leave it as an open question because I don't know
whether this will still be forced to be  a current mass. The trouble is
you have to have a formula for the lower component of $\psi$ because in the
 relativistic field theory you have to know what the lower component of a
field is.
\vskip.1in
However,
I think  you are correct. From the point of view of behaviour of
the field, since this is linear in mass, for this to be anything other than
the current mass would screw up the chiral behaviour of the field.
But what you have to be careful about is that you must not take this
expression and substitute it back into the Hamiltonian.
Once you renormalize, the Hamiltonian itself has broken its connection to
the parameters in this equation..
\vskip.1in
Let me tell you what the trouble we are having is. The second term in $(6)$
will do two things for the light-front canonical Hamiltonian:
 it will give you the $m^2$ term of the free Hamiltonian and it will
give you the linear $m$ term in the interaction Hamiltonian.
The same substitution will give you one term that has to produce a
constituent mass and another term that has to produce the current mass and
I don't know whether you can handle it except by
 breaking the connection between eqn. $(6)$ and the renormalized Hamiltonian.
You still have to have an eqn. for the lower component because you cannot
do Lorentz invariant field theory without it and you must have a
Hamiltonian but they may not be connected any more.
I want to leave it as an open question.
\vskip.1in
An easy way to say it:  you write down the canonical Hamiltonian, write
down the equation for the lower component, but the renormalization happens
separately for both. On the other hand, I use the idea that these terms can
appear here in the eqn. for the lower component of the $\psi $ field to
justify putting in the kind of terms that would have been generated
if  I could make the substitution.
\vskip.3in

{\bf Susskind:} Do you have to maintain the conservation of color except
when the coupling equals the physical coupling?

{\bf Wilson:} Susskind raises the right  question: do you conserve the color
locally? Remember, we always conserve color globally, that's trivial. But do
you have local conservation of color at any value of the coupling constant
other than the one given by the asymptotic freedom?

{\bf Susskind:} The issue is subtle because you are not separating the
tricks and artifacts of the computation and  the real physical things that
are expected to happen at the physical point.

{\bf   Wilson:}
Let's go back to what I said at the start about building the bridges:
about linking
the qualitative physical picture of the constituent quark model to
QCD through the bridge. To generate the qualitative
picture I am going to start with a weak g at the hadron mass
scale. In order to have
something that I can work with I am going to consider a coupling of
order $e^2$. I am
going to make $g^2$ just as small as $e^2$, so I can build the qualitative
picture. The bridge is the range from $g^2 =e^2$
to $g^2=g_S^2$, where $g_{S}$ is the
asymptotically free value for relativistic QCD, measured at
the hadron mass scale.
\vskip.1in
 To build the  picture at weak coupling there are some qualitative
things that I want to get. I want to get confinement. I would not tolerate a
weak coupling picture based on the Coulomb force. I've just given you the
argument why I want an explicit chiral symmetry breaking term. I will not
tolerate a situation when the $SU(3) \times SU(3)$ multiplets do not split
as I turn on the coupling. I want to have these two things. But I am
willing to sacrifice almost everything else in order to get those two
things.

So, if somebody tells me I don't have  locally conserved color current, I
say: Fine. I don't care. Because that will have to come back again as we
cross the bridge to relativistic QCD and then it should be perfectly satisfied.
\vskip.1in

{\bf Question (Susskind):} It's just that
the physical pictures that you are making may be
to some extent an artifact of the method of computation, which
continuously changes over to the physical picture when
 the coupling approaches the physical value.

{\bf Wilson:} That's correct.

{\bf Susskind:} One has to be aware of that.

{\bf Wilson:} That's right. Remember, if you are living in a house and if
you are freezing to death ....

{\bf Susskind:} I am not criticizing you!

{\bf Wilson:} I am just giving you a way to think about it.
Think about it like this: You are living in a
house and you are freezing to death, and you realize you want to save the
first floor. So you take your ax to the second and third floor, put them in
the fireplace, so that you get yourself warm. I took the ax to the
second and third stories of the QCD House
 so that I can preserve some warmth for
the first floor.

It was so devastating to try to do the QCD with the Coulomb force and no chiral
symmetry breaking ...

{\bf Susskind:} I am completely with you!

{\bf Wilson:} Right. Well, you have to show to me that it is just as bad to
break the local conservation of the color current as it was to
have no confinement and no chiral symmetry breaking
before I even worry about your problem.

{\bf Susskind:} What problem?

{\bf Wilson:}  Lack of local conservation of the color current.

{\bf Susskind:} It's not my problem.

{\bf Wilson:} Ok, fine, whoever's it was.

\vskip.3in
 {\bf Question (Berera):} You start with a bare Hamiltonian which is
defined at some large cutoff. And you want to get to an effective theory
which has some lower cutoff. Now you let the cutoff to go to infinity and
you solved the ultraviolet problem. And you hold the cutoff fixed and you
hold the parameters at your cutoff  at some typical value.

My question is:
because of including these terms
do you have any feeling
whether you are going to change the ultraviolet behaviour of the theory?
Am I going to flow to some fixed point that is not physical? There might be
technical problems with all the operators.

{\bf Wilson:} What I am expecting is that since the chiral breaking
operator reflects
a vacuum effect, not an ultraviolet effect, it will have a soft
form factor $\vec{f}_{\perp}$.
It will be something that
drops out very fast, when the $q^{\perp}$ becomes large and because it does not
affect large $q^{\perp}$ it will not affect the ultraviolet renormalization
at all. Because it is a nonlocal term, I put in a term that will directly
impact the low $q^2$ physics without affecting the ultraviolet
renormalization. Now, I may be wrong on that
in a sense that when we try
to do the calculations we may find that this $f$ (i.e. the form factor
$\vec{f}_{\perp}$)
 tries to persist at large $q$ and then I'll have a more complicated
ultraviolet renormalization,
 but at the present time I am very  optimistic and I say it does not affect
it.
\vskip.3in
{\bf Question (Susskind):} I want to see the  strategy  for going to the
physical point. And what kind of physics you expect.

{\bf Wilson:} Let me give you the strategy that I personally have in mind.
It's one of the things we are implementing in the group, not just me at
this point. The idea is if you put $g^2$ of order $e^2$ and then you
calculate the way you calculate for QED. You have nonrelativistic bound
states
just as in QED. The only difference between this and QED is that we have a
confining
potential on the top of Coulomb potential, whereas in QED you had only the
Coulomb potential. And of course in this case you have a
 nonrelativistic bound state structure
for gluons. So gluonium  exists in a whole sequence of
nonrelativistic bound states just
as quarkonium does. And it's a very orderly picture. You crank it out at
order $g^2$ just the way you do in QED. Then we borrow all the
apparatus for doing radiative corrections to QED bound states. (And you know
that the record is something like 8th order.)
So what we would do is we would calculate bound state properties
as a power series in $g$  and then
we're going to cook the data helped by
the usual round-up of power series extension methods in order to
extrapolate the power series result to the relativistic value $g_S^2$.

{\bf Question (Yan):} What about the rotational symmetry?

{\bf Wilson:}The full  rotational symmetry in x, y, z  can be expected only at
the point $g=g_{S}$.
So one of the tests of whether things work is: when you calculate everything
in the Particle Data Table, do all those spin structures of states
come into line so
that every member of a spin multiplet
has the same mass? If you can really do what the
papers of Nathan Isgur have done
where you have page after page of the tables and it
all works based on a consistent framework for extrapolating to $g_S^2$
then I will be happy.

We have lots of fudge factors  here. Every form factor
in the Hamiltonian will become a fudge
factor. It's going to be an interesting discussion when we come back and
we say we are getting the particle data table and you say: You just fudged
it. It's going to be a very interesting debate.

As for renormalization problems, we expect to learn how to get rid of all the
cutoff dependence, order by order.
All cutoff dependence will be removed. You will then not be able to challenge
us
on the cutoff dependence. You will only be able to challenge us on the
finite fudge factors, not the infinite subtractions themselves.

\vskip.3in
{\bf Susskind:} You have a computational method, I have a physical picture of
what  happens. I would like to go on record with what I'm guessing, or
speculating on the connection of the two of them and what you'll find when
you go to physical value of the coupling. Can I do that?

I will try to express it in terms of a picture, no equations.
We have  the rapidity $x$  or $\eta $ axis. As far as I
can tell your picture is something
like this:  There is the zero mode
down here, and a few modes very close to $x=0$ and we would like to
integrate them out. You put  the coupling constant small and  you
integrate them out, but you want to get the chiral symmetry breaking. So
the only way to  do this is to put some symmetry breaking associated with this
mode. The interaction feeds itself right back up to the high momentum
particle going
up here. So it is a direct interaction between what is at $x=0$ and
everywhere else on the rapidity axis.

That's alien to my picture. In my picture way it works is that the
symmetry breaking is fed up through a chain of degrees of freedom.

But, the coupling of your symmetry breaking interaction (let's call it $C(g)$)
 is a function of $g$. Now, there is some
dynamics in the system that as the coupling gets
large, the system will simply be able to support its own chiral symmetry
breaking without the direct interaction (that you put in). In other words,
the coupling $C(g)$ will go to zero at the point where $g$ goes to $g_{AS}$.

That's my guess.
\vskip.1in
{\bf Wilson:} Let me give my version of what happens.
\vskip.1in
I am going to expand $C(g)$ in powers of $g$. I expect it to look like
this:
$$ C(g) = g^2 c + g^4 \log ({p^+\over{ {\cal P} }}) +
g^6 \log ^2({p^+\over{ {\cal P}}})$$
where ${\cal P}$ is whatever you choose as a reference momentum.
 You can't have the logarithm of $p^+$
without dividing it by something.

At the 4th order my expectation is that we start getting logarithms.
Then what we'll be
arguing about is what happens to the sum of these logarithms. Before we
have the logarithms, couplings are completely
uniform all along the rapidity line. But as soon as you get
logarithms, you will start to see localization in Susskind's sense.
Then the question will be, what happens as $g \rightarrow g_S$?

\vskip.1in
And  now I say no more.

\vskip.3in
{\bf Acknowledgement}
\vskip.1in

I (K.G.W.) have been reporting work of the entire Light-Front group at OSU in
collaboration with Stan G\l azek. I am especially grateful to Daniel
Mustaki for drawing our attention to the unique features of chiral symmetry
in light-front coordinates.
\vskip.3in
{\bf References}
\vskip.1in

1. K. G. Wilson, T. S. Walhout, A. Harindranath, Wei-Min Zhang,
S. D. G\l azek and R. J. Perry, Phys. Rev. {\bf D 49}, 6720 (1994).

\end{document}